\documentclass[conference]{IEEEtran}

\makeatletter

\def\ps@IEEEtitlepagestyle{%
    \def\@oddfoot{\mycopyrightnotice}%
    \def\@evenfoot{}%
}
\def\mycopyrightnotice{%
    {\footnotesize  978-1-5386-7177-1/18/\$31.00 \textcopyright2018 IEEE\hfill}
    \gdef\mycopyrightnotice{}
}

\makeatletter
\newcommand*\titleheader[1]{\gdef\@titleheader{#1}}
\AtBeginDocument{%
  \let\st@red@title\@title
  \def\@title{%
    \bgroup\normalfont\large\centering\@titleheader\par\egroup
    \vskip1.5em\st@red@title}
}
\makeatother

\usepackage[english]{babel}
\usepackage[utf8x]{inputenc}
\usepackage[T1]{fontenc}
\def\BibTeX{{\rm B\kern-.05em{\sc i\kern-.025em b}\kern-.08em T\kern-.1667em\lower.7ex\hbox{E}\kern-.125emX}}
\usepackage{amsmath,amssymb,amsfonts} 
\usepackage{graphics}                 
\usepackage{float}
\usepackage{cases}
\usepackage[caption = false]{subfig}
\usepackage{epstopdf}
\usepackage{color}                    
\usepackage{algorithm}
\usepackage{algpseudocode}            
\usepackage{algorithmicx}
\usepackage{multirow}
\usepackage{multicol}
\usepackage{amsmath}
\usepackage[final]{graphicx}
\usepackage[colorinlistoftodos]{todonotes}
\ifCLASSINFOpdf
  
\else
 
\fi

\title{Experimental Evaluation of LoRaWAN in NS-3} 

\titleheader{\footnotesize 2018 28th International Telecommunication Networks and Applications Conference (ITNAC)} 

\begin{document}

\author{\IEEEauthorblockN{Furqan Hameed Khan$^\dag$ and Marius Portmann $^\ddag$}\\
\IEEEauthorblockA{School of ITEE, The University of Queensland, Brisbane, Australia\\
 Email: $^\dag$ furqanhameed.khan@uq.edu.au, $^\ddag$ marius@ieee.org}
}


\maketitle

 \begin{abstract}
Long Range Wide Area Networks (LoRaWAN) is a technology devised for the long range connectivity of massive number of low power network devices. This work gives an overview of the key aspects of LoRaWAN technology and presents results that we achieved via extensive evaluation of Class A LoRaWAN devices in different network settings using the state-of-the-art network simulator (NS-3). At first, we focus on a single device and its mobility. We further undertook evaluations in an extended network scenario with a changing number of devices and traffic intensity. In particular, we evaluate the packet delivery ratio (PDR), uplink (UL) throughput, and sub-band utilization for the confirmed and unconfirmed UL transmissions in different environments. Our results give new insights for future efforts to optimize the LoRaWAN performance for different large scale Internet of Things (IoT) applications with low power end devices.
\end{abstract}


\IEEEpeerreviewmaketitle

\section{Introduction}
In recent years, several technologies have been developed to achieve sustainable coverage for large scale Internet of Things (IoT) devices that will become an integral part of future smart cities. These technologies, featuring low power and long range connectivity  to IoT in the licensed as well as the unlicensed spectrum. In the licensed band, LTE-M and narrowband IoT (NBIoT)~\cite{IEEEhowto:01} are cellular IoT technologies that achieve reliable connectivity for a massive number of devices. While licensed operation of IoT devices is costly, it can achieve better quality of service (QoS)~\cite{IEEEhowto:00}. In unlicensed bands, SigFox~\cite{IEEEhowto:000} and LoRa~\cite{IEEEhowto:07} are two proprietary technologies that are increasingly deployed in different IoT applications. SigFox uses ultra narrowband channels ($\approx$ 100 Hz) within the 869 MHz (Europe) and 915 MHz (US) bands~\cite{IEEEhowto:02}. It is suitable for applications that require extremely low uplink (UL) rates with a maximum payload size of 12 bytes. LoRaWAN, on the other hand uses the LoRa modulation scheme over its PHY layer and achieve a better UL rate, ranging from 0.3kbps to 27kbps~\cite{IEEEhowto:03}.

\begin{figure}[!t]
\begin{center}
\resizebox{3.2 in}{2.0 in}{\includegraphics{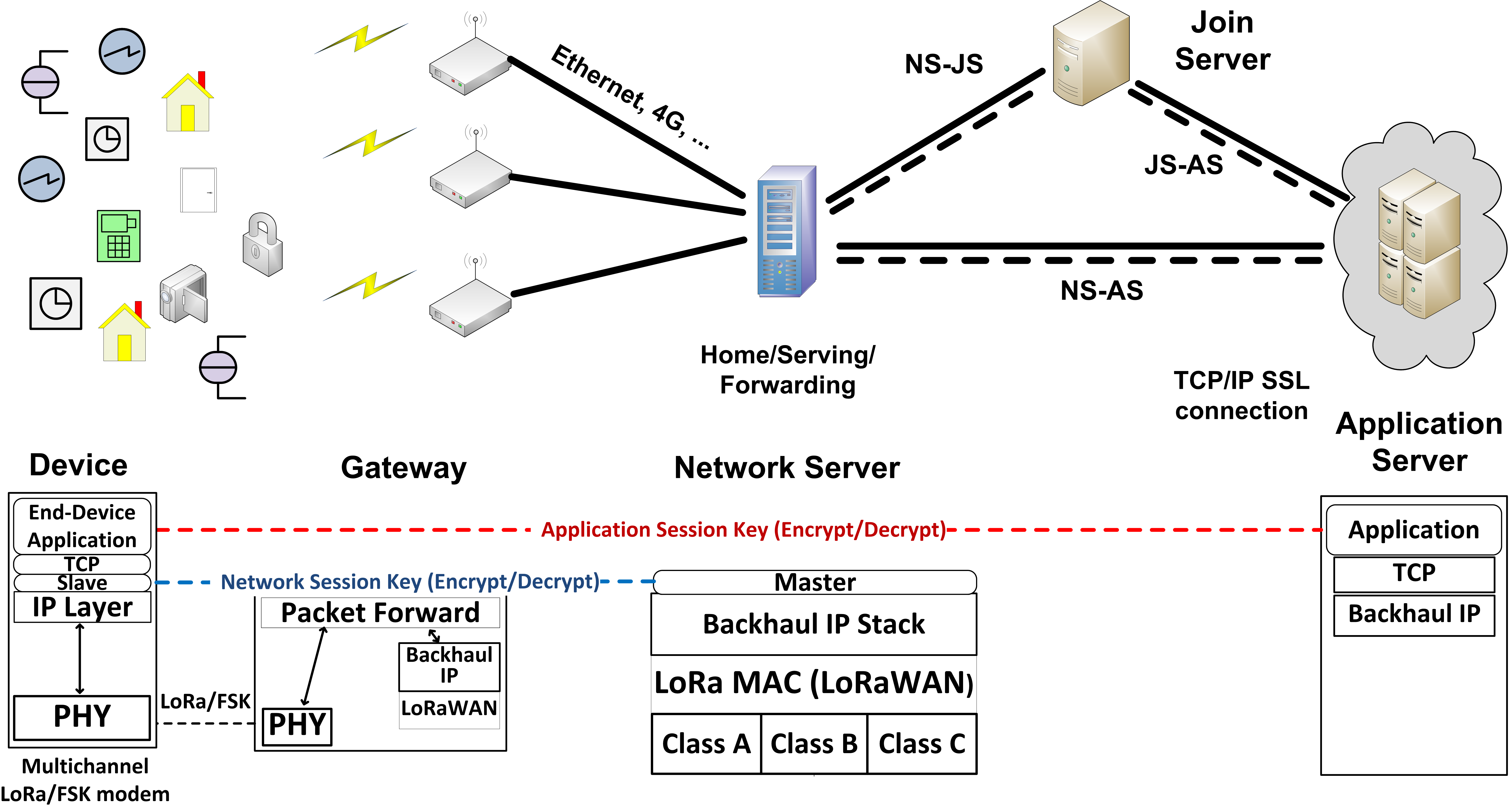}}
\caption{\em LoRaWAN Network Architecture and Network Component Layers~\cite{IEEEhowto:08}}
\label{jpg1}
\end{center}
\end{figure}

Fig.~\ref{jpg1} shows a typical LoRaWAN architecture consisting of devices connected to the network server through one or multiple gateways. The interface between the server and gateway can be Ethernet, 3G/4G, etc. The LoRaWAN architecture is designed to achieve optimal operation for constrained application (CoAP) environments. Devices are connected over-the-air to LoRa gateways and are served through unlicensed region specific ISM bands defined in~\cite{IEEEhowto:05}. To send packets over the network, a device can either be activated over-the-air or it can be manually registered using activation by personalization (ABP) at the network server before its deployment~\cite{IEEEhowto:04}. The gateways are just traffic forwarding elements and implement the physical layer processing of LoRaWAN packets. As shown in Fig.~\ref{jpg1}, the LoRa PHY performs LoRa/FSK modulation/demodulation of the DL/UL packets and forwards them to the corresponding device/network server, respectively. Typically over the UL, the device performs the application specific information processing and forwards packets to the NS via one or multiple gateways. The LoRa medium access control (MAC) layer management at NS includes removing duplicate UL messages, processing acknowledgements (ACKs) for confirmed frames, and assigning MAC layer parameters e.g. channel, data rate/spreading factors (DR/SF), etc. to different network devices. The join server (JS), as shown in Fig.~\ref{jpg1} keeps track of all network session keys and application session keys issued to network devices as well as the network server and application server associated with that device. This enables secure data transfer between session establishment and termination procedures.

Due to low deployment cost, simple network maintenance, energy efficiency, and higher achievable rates, LoRaWAN is getting increasingly adopted for different industrial applications, ranging from agriculture and environment monitoring, to various smart city applications~\cite{IEEEhowto:03}. With this increasing popularity, it is important to explore its operation under different settings and use cases. The current LoRaWAN specification~\cite{IEEEhowto:04} does not provide details about various aspects of its operation that may affect the resulting performance gains. Therefore an in-depth study of different operational features of LoRaWAN is required using a physical testbed or a simulation tool.

This paper presents an experimental study of key performance metrics in a LoRaWAN network using system-level simulations. We evaluate parameters such as channel utilization and packet delivery ratio (PDR) for confirmed and unconfirmed UL transmissions in different scenarios. Our study reveals new key insights for service providers on how LoRaWAN efficiency is affected by the variation in network parameters e.g. spreading factor (SF), number of transmissions, channels, and the network load.
%

\section{Background}
\subsection{Layers in LoRaWAN}
The physical (PHY) layer in LoRaWAN makes use of LoRa, which is a chirp-spread spectrum (CSS) modulation scheme that uses a chip-coded sequence to spread the data-sequence signal over a much wider bandwidth~\cite{IEEEhowto:031}. This makes the modulated signal robust against noise and multi-path fading. The higher the chip rate ($DR_{chip}$) compared to the original data rate ($DR$), the wider the frequency range over which the transmitted data spreads and the longer the distance it can reach. The log-ratio of the chip rate to the original data sequence rate is called the spreading factor (SF), defined as $SF$ = $log_2$($DR_{chip}$/$DR$)~\cite{IEEEhowto:03,IEEEhowto:031}. Different SFs are defined (from 7 to 12) based on the rate of the chip sequence used. Note that the chip sequences for different SFs are orthogonal to each other, hence giving additional diversity in the achievable channel gain. In short, LoRa in itself hides the LoRaWAN complexity and enables different SF configurations over a channel with a variety of bandwidths (e.g. 125 kHz, 250 kHz, and 500 kHz) and code rate settings.

The medium access control (MAC) layer uses LoRaWAN that supports the operation of different types of LoRa devices namely, Class A, B, and C. Class A devices enable energy efficient data frame transmission only in the UL. All LoRaWAN devices must support UL operation which makes this class as the baseline class of LoRaWAN. The downlink (DL) transmissions for Class A devices must follow an UL data frame sent by an end device application. For this reason, the LoRaWAN frame header includes a pending data field, which is set if there is more DL data needed to be sent to the device. Based on this field, a device wakes up to receive the next DL data frame. Class B devices focus on the DL transmission of data frames. This class support extra slots known as $ping$ $slots$ that can be scheduled by the NS within each beacon duration of 128 seconds. Note that in both of these classes (A and B), the UL and DL operations are suitable for devices with limited energy supply. Class C type devices remain in the active state all the time to receive data from the NS as they have no energy constraints.
\subsection{LoRaWAN Operation}
\begin{figure*}[t!]
\begin{center}
\resizebox{4.0 in}{3.0 in}{\includegraphics{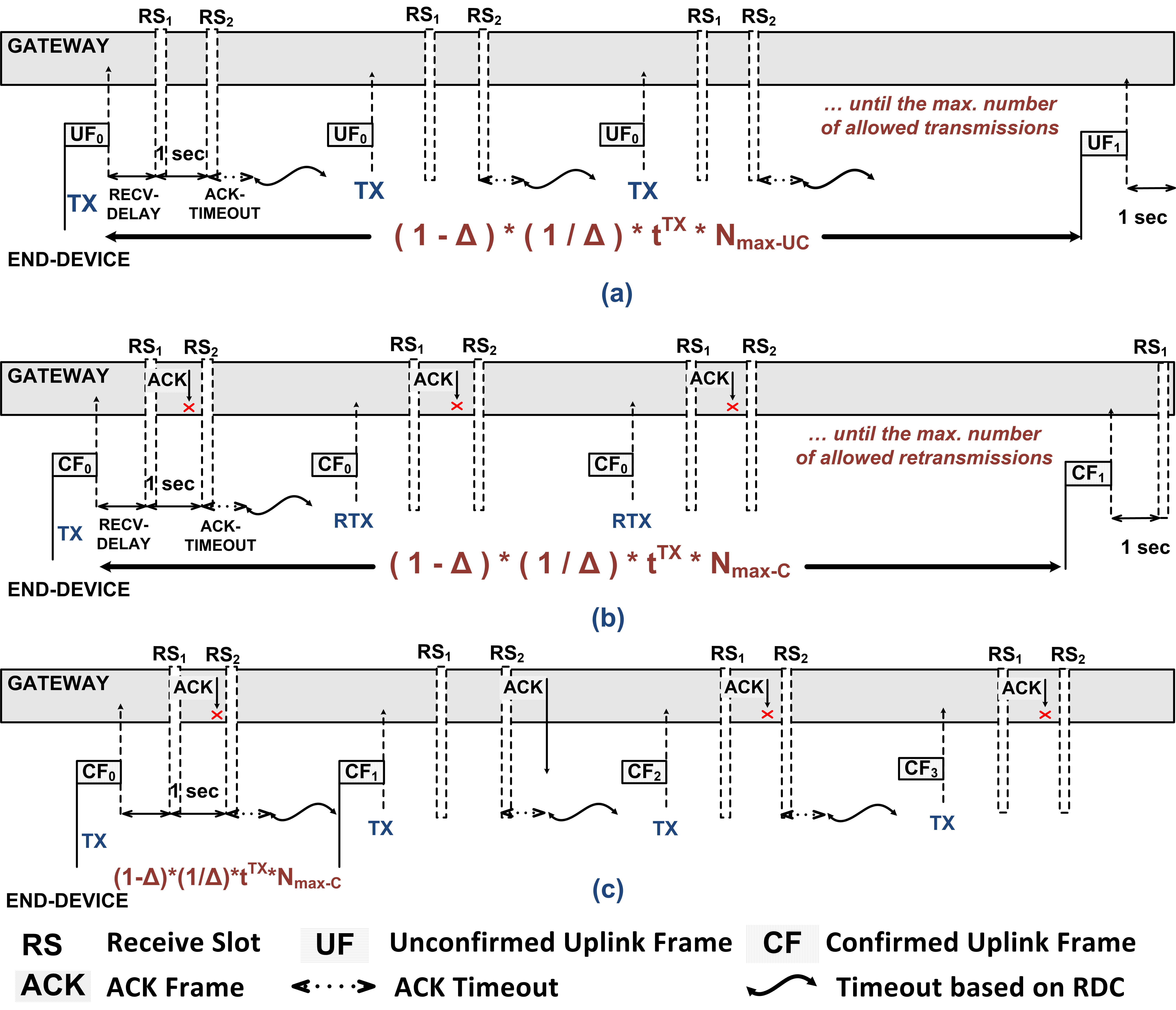}}
\caption{\em UL transmission scenario with different maximum tx. limits in LoRaWAN: When (a) $N_{max-UC}$ = 4, (b) $N_{max-C}$ = 8, (c) $N_{max-C}$ = 1}
\label{jpg2}
\end{center}
\end{figure*}

As described earlier, any UL transmission from a class A type device requires it to open two receive slots on different channels, one after the other. As shown in Fig.~\ref{jpg2}a, initially the device will open the first slot ($RS_1$) after the time length of RECV-DELAY~\footnote{From LoRaWANv1.1, recommended value of RECV-DELAY is 1 second.}, and then it will open the slot 2 ($RS_2$) exactly after one second of $RS_1$. The LoRaWANv1.1 protocol supports two types of UL transmissions from Class A devices to the gateway, an UL unconfirmed transmission and a confirmed transmission. An unconfirmed UL transmissions does not require an ACK from the gateway as shown in Fig.~\ref{jpg2}a. In contrast to unconfirmed transmissions, a confirmed UL transmission requires the gateway to send an ACK frame if the frame is received successfully. An ACK frame can be sent in either one of the two receive slots, that are opened following an UL data frame transmission, as shown in Fig.~\ref{jpg2}b. Each UL transmission from the LoRaWAN device must follow the regulatory duty cycle (RDC) constraint which is specifically defined for different available sub-bands and regions in~\cite{IEEEhowto:05}. The LoRaWAN MAC header defines a frame counter field within the UL and DL frame that counts the number of frames sent over the UL/DL, respectively. For each next UL and DL frame the counter gets incremented by one. The maximum number of retransmissions limits the lora-device MAC (e.g. LoRaWAN) to retransmit the confirmed UL frame up to maximum $N_{max-C}-1$ times, in case no ACK is received. Note that maximum transmission $N_{max-C}$ can be restricted to 1, resulting for a device to move to the next frame immediately after an RDC timeout following an UL transmission as shown in Fig.~\ref{jpg2}c. The LoRaWAN specification~\cite{IEEEhowto:04} recommends the maximum number of retransmissions as 7~\footnote{e.g. including the first transmission, overall $N_{max-C}$ will be 8.}. Also note that each retransmission is done on a different channel and the SF for each second confirmed frame retransmission is increased~\cite{IEEEhowto:16}.
%
\section{Related Work}
Previous research on LoRaWAN broadly used either physical test-bed experiments,  network simulation based evaluations, or mathematical modeling for LoRaWAN network study.

In~\cite{IEEEhowto:15}, the authors studied the received signal strength indicator (RSSI) and coverage using different settings of SFs on a physical LoRa test-bed consisting of a single device and gateway. The results compare the measured RSSI with the one specified in the LoRa device data sheets and show that the actual RSSI is around 6 dB less than the one specified in the data sheets. Furthermore, the field tests show that the LoRa gateway can achieve a range of up to 3km in the suburban environment.

Other works such as ~\cite{IEEEhowto:02,IEEEhowto:14,IEEEhowto:09} used simulation tools to study various aspects of LoRaWAN. In~\cite{IEEEhowto:02,IEEEhowto:14}, the authors used LoRaSim~\footnote{http://www.lancaster.ac.uk/scc/sites/lora/}, for the scalability analysis of LoRaWAN in different environments.~\cite{IEEEhowto:02} investigates the data extraction rate and the overall network energy consumption in a network with a single and multiple sinks. The results show that LoRa enables scalability, provided that different transmission parameters (e.g. channels, SFs, etc.) are carefully chosen. Similarly, the authors in~\cite{IEEEhowto:14} did experiments to analyze the effect of DL traffic over the UL throughput performance in LoRaWAN. In this regard, the work shows that even in the presence of a small fraction of DL traffic, achieving scalability is crucial as the UL goodput decreases sharply.~\cite{IEEEhowto:14} takes a step further and emphasizes that it is important to efficiently choose the number of gateways and retransmissions to support the required UL performance for the end nodes. Next, ~\cite{IEEEhowto:09} introduces a new LoRaWAN network simulation module built in NS-3. Using different network settings, the work evaluates the PDR over the UL and DL confirmed/unconfirmed messages where all end devices and gateways are using a single UL channel (868.1 MHz) other than the high power reserved channel for $RS_2$. The results show that the PDR for confirmed frames is highly affected, especially when the transmission period is small. Next, following the works of~\cite{IEEEhowto:02,IEEEhowto:14}, the NS-3 based simulation experiments of LoRaWAN in~\cite{IEEEhowto:09} also reaffirms the finding that the UL PDR gets highly affected in the presence of DL traffic.

Further studies~\cite{IEEEhowto:03,IEEEhowto:16} have used a Pure Aloha based mathematical model to explore the performance limits of LoRaWAN. In~\cite{IEEEhowto:03}, the authors perform an analysis of performance achieved using LoRaWAN in terms of the number of successfully received packets per hour. The works show that when the packet transmission rate is low, the UL throughput is mainly limited by the number of collisions. In contrast, when the transmissions rate is high, the duty cycle limitation limits the network performance. As such, the study suggests that further modifications should be made in LoRaWAN to enable support for different use case scenarios. The study also gives some new research directions, e.g. designing new channel hopping schemes, multi-hop transmissions in LoRaWAN, and TDMA based UL scheduler over LoRaWAN. Similarly, the work of~\cite{IEEEhowto:16} proposes a mathematical model of LoRaWAN channel access. The model can be used to estimate the packet error rates considering the capture effect for a given range of network traffic load.

In this work, we explore LoRaWAN using the NS-3 based LoRaWAN module (discussed in~\cite{IEEEhowto:09}). NS-3 is one of the most widely used network simulation platform and it allows us to explore various performance metrics in different network settings. This paper presents new results illustrating the effect of the maximum transmission limits ($N_{max-C}$/$N_{max-UC}$) and the number of available channels on the resulting PDR and sub-band utilization, in the case of confirmed and unconfirmed UL transmissions.

The next section gives details about our experimental methodology. This includes a description of the NS-3 based LoRaWAN module, different performance metrics, and the simulation environment of our experiments. Section V gives details of each use case and provides the corresponding results from our experiments using the NS-3 based LoRaWAN module. Later in Section V, the simulation results are discussed showing PDR, UL throughput, sub-band utilization as well as the impact of network load and the sub-band resources on the resulting performance. Finally, section VI concludes this study.
\section{Experimental Methodology}
A reasonably sized LoRaWAN network requires the deployment of several low power sensor devices that may scale up to several thousand nodes.  Building and maintaining such a huge network is not trivial, expensive, and time consuming. Therefore, a common way is to use simulation to provide an approximation of a real network environment.

\subsection{LoRaWAN Module in NS-3}
\begin{figure}[t!]
\begin{center}
\resizebox{3.2 in}{2.5 in}{\includegraphics{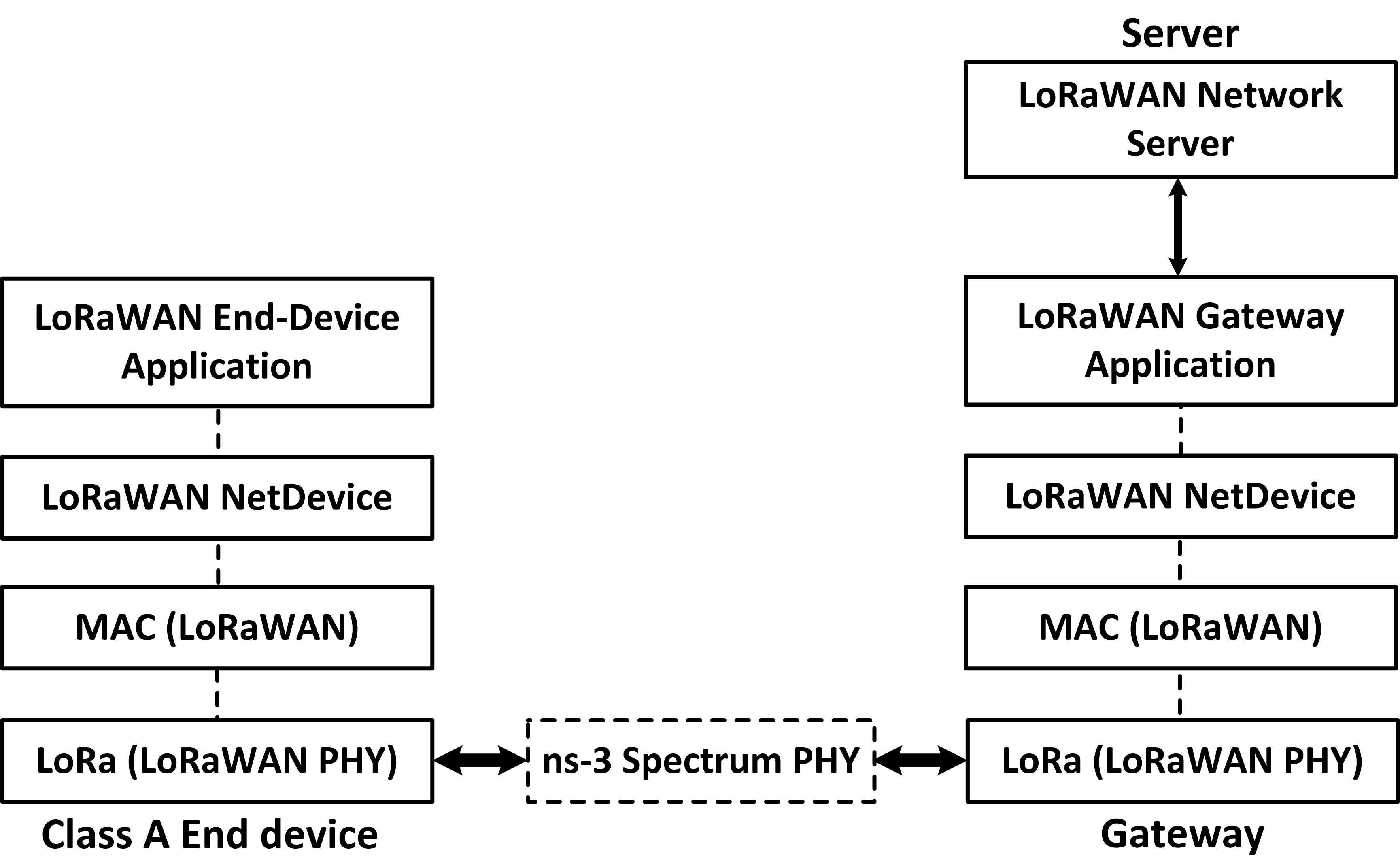}}
\caption{\em LoRaWAN module setup in NS-3: Class A LoRaWAN devices, gateway, and the NS~\cite{IEEEhowto:09}.}
\label{jpg3}
\end{center}
\end{figure}

The NS-3 LoRaWAN module is an extension of the NS-3 module for the low power wide area network (LPWAN)~\cite{IEEEhowto:11}. The LoRaWAN MAC/PHY components are running over each LoRa device and the gateway. The PHY layer of each device interacts with that of the respective gateway via the NS-3 Spectrum PHY module, implementing the devices air interface and channel specific parameters, as shown in Fig.~\ref{jpg3}. As mentioned earlier, on each channel in a sub-band, a gateway can receive signals with different SFs simultaneously. Note that in the testbed the LoRa PHY uses the error model from the baseband implementation of the PHY layer in MATLAB, simulations considering an AWGN channel as described in~\cite{IEEEhowto:09}. The collision model used in the NS-3 module is based on the capture effect. The capture effect comes into play when during the collision of two simultaneous UL transmissions (with same frequency and SF) the stronger signal captures the weaker signal. As a result, the frame with stronger received power is successfully received by the gateway while the frame with a weaker receive power is lost.

In the LoRaWAN module, for the case of confirmed UL transmissions, the choice of receive slots ($RS_1$/$RS_2$) by the gateway for sending ACKs back to the device is based on the following:
\begin{itemize}
\item RDC restriction over the sub-band.
\item Over the given channel and SF the LoRaWAN MAC is in idle state.
\item No other MAC data is scheduled on the particular channel with same bandwidth and SF.
\end{itemize}
\subsection{Performance Metrics}
We consider a LoRaWAN network with a NS, one gateway, and $|A|$ number of devices in the network. A device $a$ ($\in$ $A$) application generates packets following Poisson law with average rate $\lambda$, regarded as the traffic arrival rate. We assume this packet generation rate is fix for all network devices. Next, all packets generated by the device application are forwarded to the MAC layer to send immediately over the UL only if the RDC timer at the device from earlier transmission has expired and the device MAC is in idle state. 

In our system, we assume only Class A UL frame transmissions from the end devices. The DL frames contain ACKs and are only sent in case of confirmed UL transmissions. If $t^{TX}$ is the air time of transmitted data frames, then following the RDC constraint ($\Delta$), a device can transmit data after a period of at least $\frac{t^{TX}}{\Delta}$ seconds. In other words, the maximum packet transmission rate is $\frac{\Delta}{t^{TX}}$. 

Next, as shown in Fig.~\ref{jpg2}, we use $N_{max-UC}$/$N_{max-C}$ to represent the maximum number of UL unconfirmed/confirmed frames transmissions respectively, from an end device. 
\begin{table}[t!]
\caption{Symbols and their definition}\centering
\label{table2:parameters}
\begin{tabular}{|c|c|}
\hline
Symbols  & Definition \\
\hline
$T$ & Overall simulation duration \\
\hline
$t^{TX}$ &  Data frame transmission time \\
\hline
$\Delta$ & UL duty cycle \\
\hline
$N_{max-UC}$/$N_{max-C}$ & Max. frame transmissions \\
\hline
$X_0$/$X_1$ & Actual tx. for unconfirmed/confirmed frame \\
\hline
$s^{x}_a$/$r^{x}_a$ & UL sending/receiving rate  \\
\hline
$\rho$ & Sub-band load \\
\hline
\end{tabular}
\end{table}
For a typical UL confirmed transmission, we assume $f_R(1,N_{max-C})$ is a path loss and load-dependent function that gives a random integer between 1 and $N_{max-C}$ as an output in the number of transmissions required for the confirmed frame reception. If $X_x$ defines a function that gives the number of transmissions of unconfirmed/confirmed frames when $x$ is 0 or 1, respectively. We can state this for unconfirmed/confirmed UL transmission as,
\begin{subnumcases}{X_x =}
N_{max-UC}, & for $x = 0$\\
f_R (1, N_{max-C}), & for $x = 1$
\label{eq01}
\end{subnumcases}
We consider one sub-band having $m_c$ channels is available to all the LoRaWAN devices for UL transmission. Following the RDC constraint of a sub-band with traffic arrival rate $\lambda$, the average frame generation rate of a device $a$ from Fig.~\ref{jpg2} is given as:
\begin{subnumcases}{s^{x}_a =}
\frac{1}{(\frac{1}{\Delta}) \cdot t^{TX}}, & if $\lambda$ $\geq$ $\frac{1}{(\frac{1}{\Delta}) \cdot t^{TX} \cdot X_x}$\\
\lambda \cdot X_x, & if $\lambda$ $<$ $\frac{1}{(\frac{1}{\Delta}) \cdot t^{TX} \cdot X_x}$
\label{eq02}
\end{subnumcases}
Here, $\Delta$ is the RDC limit over the maximum sub-band airtime that can be used by a device and $(\frac{1}{\Delta}) \cdot t^{TX}$ is the RDC period~\footnote{average period in which a device can send at most one UL frame.} of the sub-band. 

Using $\lambda$ and the RDC $\Delta$, we define the traffic intensity $t_I$ for a device $a$ as the ratio of average packet generation rate by device application ($\lambda$) to the maximum frame transmission rate $(\frac{\Delta}{t^{TX}})$, e.g. $t_I$ = $\lambda$ $\cdot$ $(\frac{1}{\Delta}) \cdot t^{TX}$. Note that since $\lambda$ is fixed for all $|A|$ devices, $t_I$ will be also be same for all network devices ($\forall$a $\in$ A).

In a sub-band with $m_c$ channels, the load $\rho$ over the time duration (0, T] in the presence of $|A|$ devices is,
\begin{equation}\label{eq03}
	\rho = \frac{t^{TX}}{m_c} \cdot \sum_{a=1}^{|A|} s^{x}_a,
\end{equation}
%
%
The above equation defines the fraction of the sub-band (with $m_c$ available channels) air-time used by all $|A|$ devices for UL data transmission. Let $r^{x}_{a}$ be the rate (of type $x$, i.e. confirmed or unconfirmed) of successfully received UL frames from a device $a$ over a sub-band. Then the capacity utilization of the sub-band (e.g. $U$) is: 
\begin{equation}\label{eq04}
	U = \frac{t^{TX}}{m_c} \cdot \sum_{a=1}^{|A|} r^{x}_{a},
\end{equation}
In other words, the utilization in above equation is the ratio of total air-time of successfully received UL frames to the overall time $m_c$ channels in the sub-band are available over UL. It is important to note that, since the utilization is specific to the given sub-band, therefore it is calculated at the MAC (LoRaWAN) layer. 

We next define the PDR as the ratio of successfully received application packets to the number of packets transmitted by the source application. It can be mathematically stated for the considered LoRaWAN network as:
\begin{equation}\label{eq05}
	PDR = \frac{1}{|A|} \cdot \sum_{a=1}^{|A|} (r^{x}_{a, App}/s^{x}_{a, App}).
\end{equation}
%
\subsection{Simulation Parameters}
The parameters used in our experiments are listed in Table.~\ref{table1:parameters}. Furthermore, our results are based on the following assumptions:
\begin{itemize}
 \item Only the UL traffic from Class A devices is considered, which can consist of either unconfirmed or confirmed frames.
 \item All UL transmissions from a device follow the specifications of EU863-870~\cite{IEEEhowto:05} e.g. UL transmissions follow 1\% RDC (as defined in Table.~\ref{table1:parameters}), whereas DL transmissions adhere to a RDC of 10\%.
 \item Each device $a$ $\in$ $A$ is sending equally sized frames over the UL with traffic intensity $t_I$.
\end{itemize}
\begin{table}[t!]
\caption{Simulation Parameters}\centering
\label{table1:parameters}
\begin{tabular}{|c|c|}
\hline
Simulation Parameters   & Value \\
\hline
UL device tx. power &  14dBm \\
\hline
Gateway coverage radius & Scenario I (7km), II (5km) \\
\hline
Spreading factor (SF) & SF12 \\
\hline
Preamble length & 8 bytes \\
\hline
Frame PHY payload & 21 bytes \\
\hline
ACK payload & 12 bytes \\
\hline
Code rate  & $\frac{4}{7}$ \\
\hline
$\Delta$ & 1\% (for UL) \\
\hline
Channel bandwidth & 125 kHz \\
\hline
Path loss model & LogDistancePropagationLoss \\
\hline
\end{tabular}
\end{table}
\section{Simulation Scenarios and Results}
This section defines the simulation environment and the use case scenarios for the evaluation of different performance metrics as explained above.
 \begin{figure}[!ht]
 \begin{center}
 \resizebox{3.4 in}{1.80 in}{\includegraphics{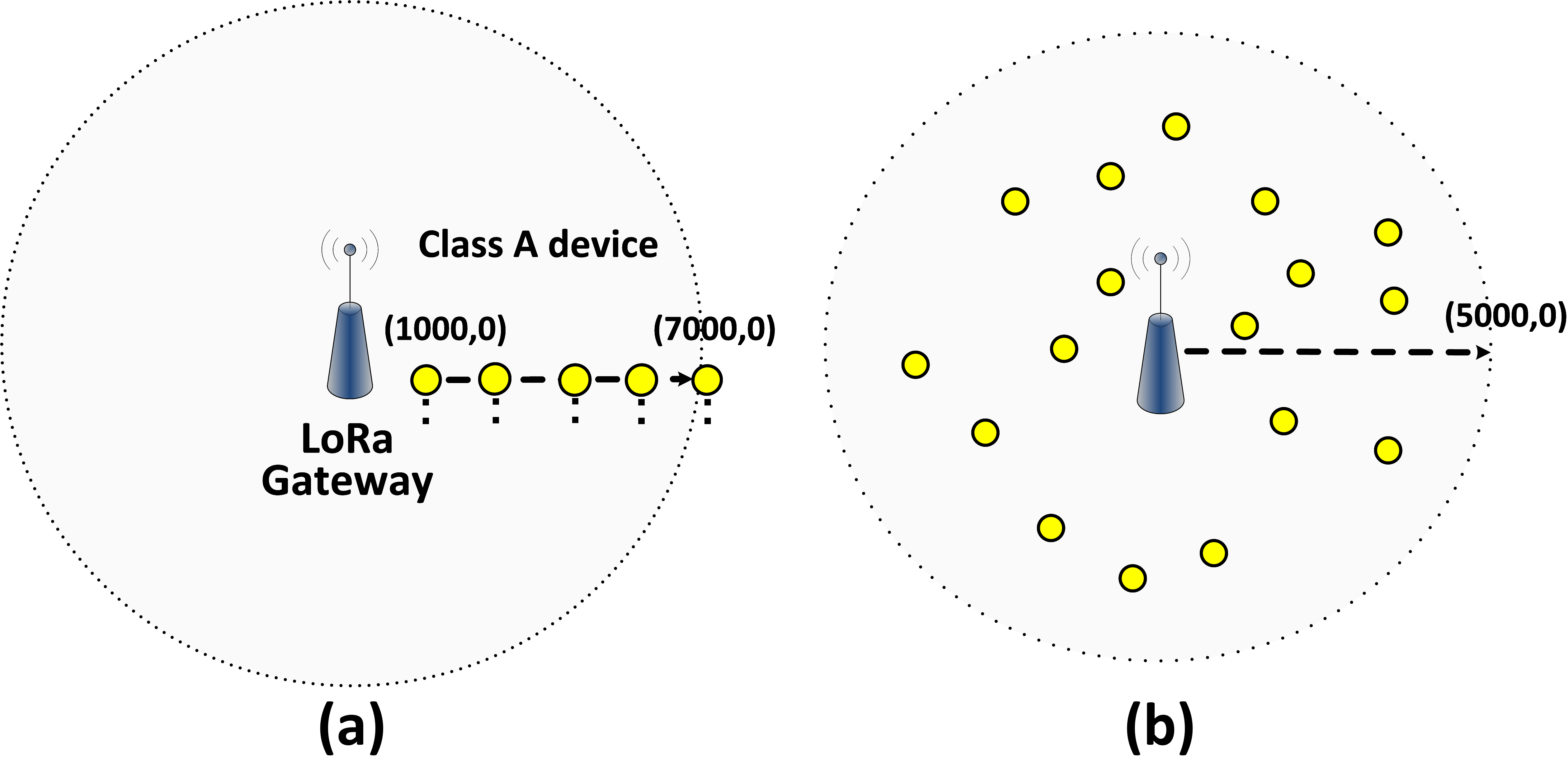}}
 \caption{\em (a) Scenario I: Single device mobility, (b) Scenario II: Multiple devices uniformly distributed over the 5km radius.}
 \label{jpg4}
 \end{center}
 \end{figure}
\subsection{Scenario I - Single Device}
We consider a single gateway network connecting end devices to the NS. The server can manage the channel and SFs for a device through the gateway to achieve better performance. Initially, the network is serving a single device and one sub-band (with only one channel) is available for the UL data transmission. The end device is allocated a fixed SF by the server.
\begin{itemize}
\item Single gateway single device: As an initial case, an end device moves along a straight line, i.e. from distance of 1 km from the gateway, to 7 km away from the gateway (as shown in Fig.~\ref{jpg4}a). After some fixed time, the node sends 100 packets over the UL. Here, we are interested in evaluating the impact of mobility, or more precisely distance from the gateway, on the PDR and the UL throughput. Note that here we consider only unconfirmed frame transmission over the UL.
\end{itemize}
\subsection{Results for Single Device Mobility}
\subsubsection{Packet Delivery Ratio (PDR)}
\begin{figure}[!ht]
 \begin{center}
 \resizebox{3.4 in}{2.0 in}{\includegraphics{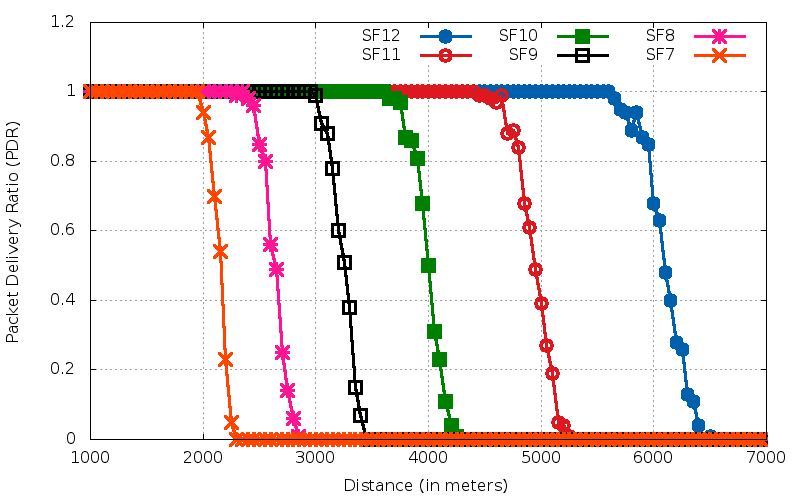}}
 \caption{\em UL PDR from a single mobile device with fixed SF.}
 \label{jpg6}
 \end{center}
 \end{figure}

Fig.~\ref{jpg6} plots the resulting PDR for the different fixed SFs setup over the given channel as the device moves away from the gateway. It can be seen that an UL rate setting of SF7 limits the communication range of the gateway to around 2km, and after $\approx$ 2.2km the UL frame PDR reduces to zero, as all frames are lost due to the low signal to noise ratio (SNR) at the receiver. Note that a higher SF increases the receiver sensitivity and enables better coverage and range over a larger distance. As a result, the UL PDR for SF12 remains close to 100\%, even though the device is $\approx$ 6km away from the gateway.
\subsubsection{Uplink (UL) Throughput}
\begin{figure}[!ht]
\begin{center}
\resizebox{3.4 in}{2.0 in}{\includegraphics{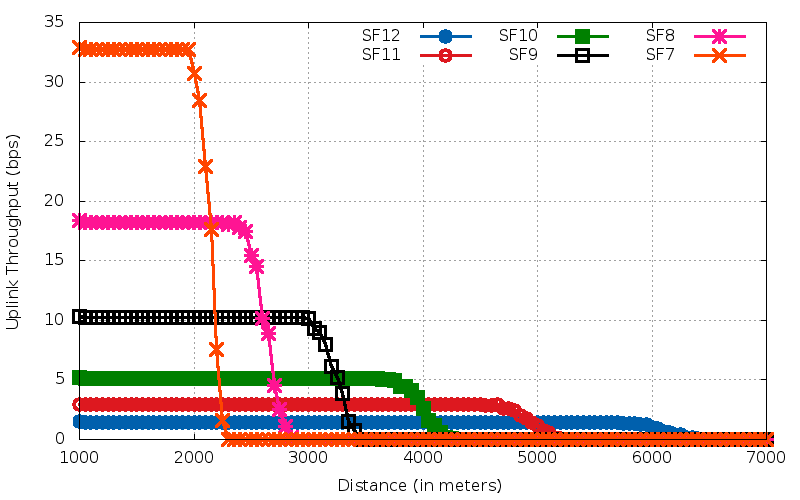}}
\caption{\em Uplink (UL) throughput for the mobile device with different fixed SF.}
\label{jpg7}
\end{center}
\end{figure}

Next in Fig.~\ref{jpg7}, the result shows the UL throughput (bits per second) for a fixed SF. It can be seen that with low SF (e.g. SF7) a higher UL throughput can be achieved. However, it will only last for some small distance, as shown in Fig.~\ref{jpg7}, due to low receivers sensitivity. Similarly, with higher SF (e.g. SF12), all UL frames up to a distance of $\approx$ 6km can be successfully received. 

It is important to note that based on the above results, an optimal scheme can be devised that adaptively chooses the SFs for different devices to maximize their UL rates. By using an adaptive rate selection, it is expected that a mobile device can achieve a better overall UL rate compared to a fixed SF scheme.
\subsection{Scenario II - Multiple Devices, Poisson traffic pattern}
In this scenario, we consider a single gateway network with multiple devices uniformly distributed over a circular area with 5km radius, as shown in Fig.~\ref{jpg4}b. All devices are configured to send UL data with SF12. For frame collisions, the model implemented in the LoRaWAN NS-3 module is based on the capture effect. Furthermore, where more than one channel is available over the sub-band (i.e. $m_c$>1), we assume that all channels have an equal probability to be selected by the device for each new UL transmission.

\begin{figure}[!ht]
\begin{center}
\resizebox{3.0 in}{2.0 in}{\includegraphics{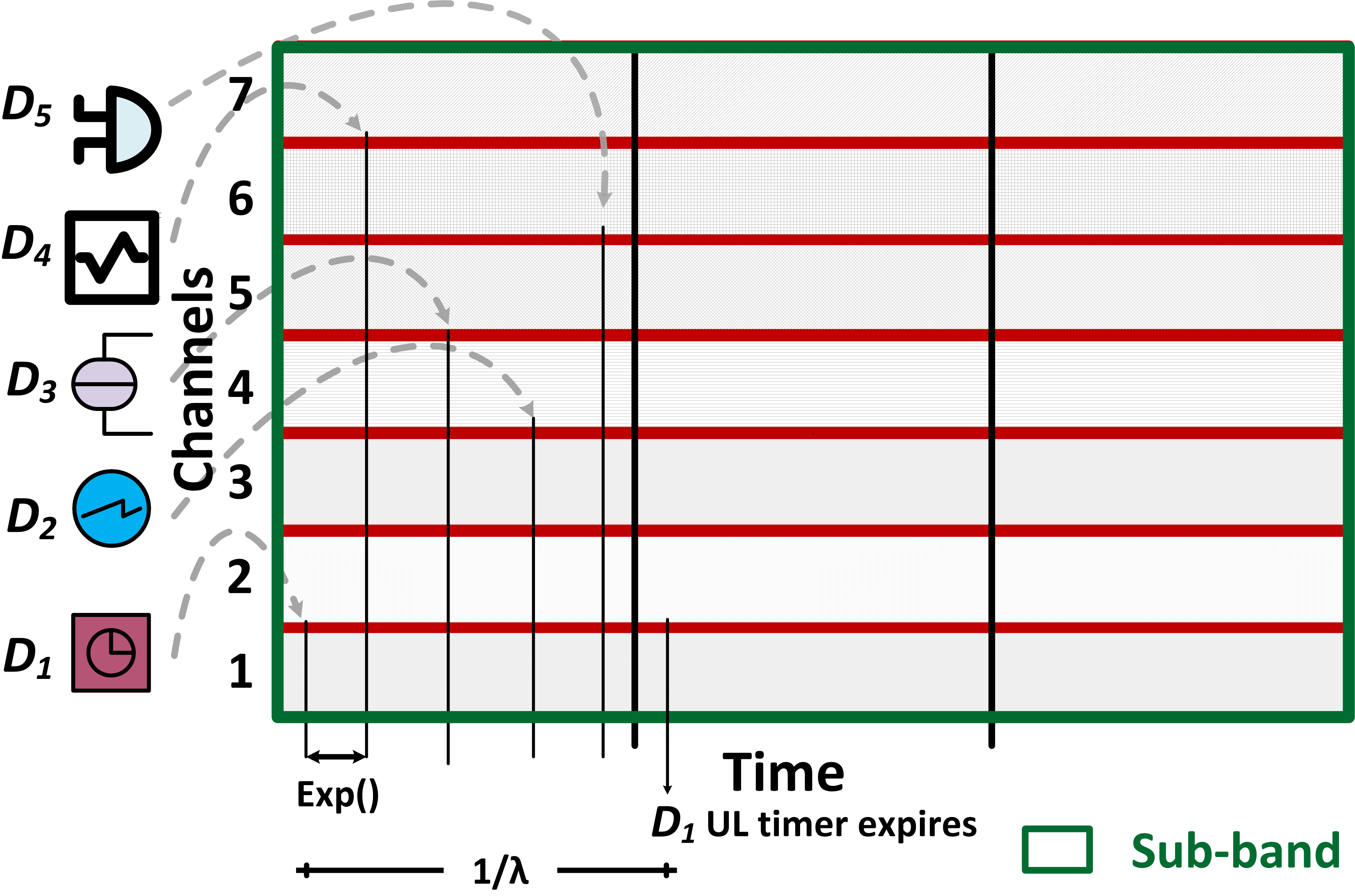}}
\caption{\em  Poisson arrivals over the sub-band with average rate $\lambda$}
\label{jpg5}
\end{center}
\end{figure}

At first, we consider a single channel gateway network (i.e. a sub-band with only one available channel) with traffic arrival following a Poisson process with exponentially distributed mean inter-arrival time $\frac{1}{\lambda}$, as shown in Fig.~\ref{jpg5}. The impact of the increasing load (number of devices) upon the network performance metrics such as sub-band utilization and PDR, is evaluated separately for the cases of confirmed and unconfirmed UL transmissions.
\subsection{Results for Multiple Devices}
In the following, we describe the results of scenario II.
\subsubsection{Sub-band Utilization ($U$)}
Considering different traffic intensities ($t_I$) as well as maximum transmissions for confirmed and unconfirmed UL frames (e.g. $N_{max-C}$ and $N_{max-UC}$), this section evaluates the sub-band utilization results.
\begin{itemize}
\item \textbf{Varying transmissions ($N_{max-UC}$/$N_{max-C}$)}
The utilization results in Fig.~\ref{jpg8b} are achieved in a fully loaded (e.g. $t_I$=1) network environment. Note that with $N_{max-UC}$=1, the utilization gets higher in Fig.~\ref{jpg8b} as the UL transmissions are able to fully utilize the capacity as the number of devices grows. Intuitively, it can also be seen that the utilization results for confirmed UL transmissions with different $N_{max-C}$ does not change as expected. This is because the ACK transmission from the gateway over the same channel increases collisions and decreases the number of successfully delivered frames. As a result the resource utilization in case of confirmed frames is considerably lower than for unconfirmed ones. Next, when maximum transmissions increase from 1 to 8 (i.e. $N_{max-UC}$=8) the utilization remains unaffected. This is because the channel is used up to its maximum limit in both cases, and therefore the collision probability does not change. From an application perspective, the utilization for the unconfirmed case will be small when maximum number of transmissions is high, even though multiple copies of same frame are received successfully. However, only one of them will be forwarded to the application, and counts as a received packet at the application layer.
\begin{figure}[!ht]
\begin{center}
\resizebox{3.4 in}{2.0 in}{\includegraphics{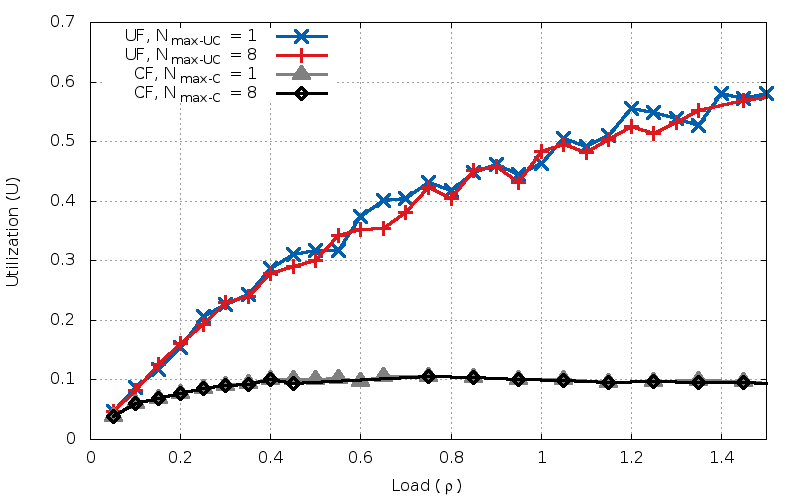}}
\caption{\em Sub-band utilization vs. network load with varying maximum number of allowed confirmed/unconfirmed UL frame transmissions with fixed SF12, $t_I$=1.0, $m_c$=1.}
\label{jpg8b}
\end{center}
\end{figure}
\item \textbf{Varying traffic intensity ($t_I$)}
The results in Fig.~\ref{jpg9} show the impact of traffic intensity ($t_I$) on the sub-band utilization for confirmed and unconfirmed UL transmissions. As expected, when the traffic load is high, the utilization also increases for both unconfirmed/confirmed UL transmissions. It is important to observe in Fig.~\ref{jpg9} that in a less loaded environment, unconfirmed transmission still achieves better channel utilization.
\begin{figure}[!ht]
\begin{center}
\resizebox{3.4 in}{2.0 in}{\includegraphics{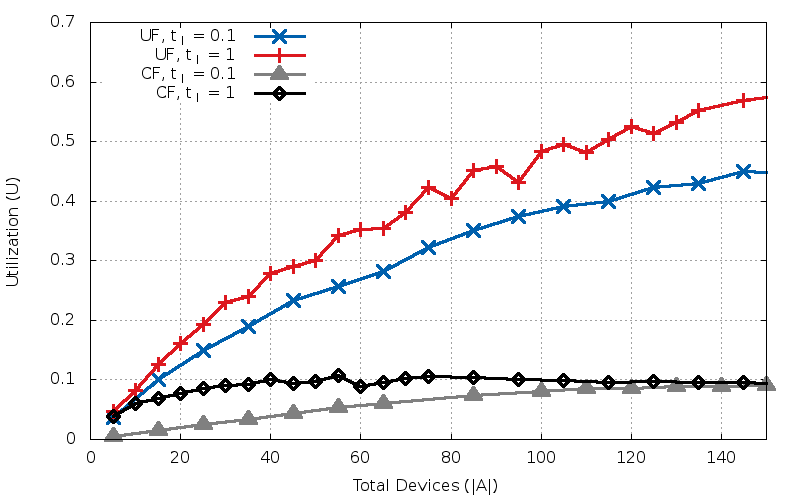}}
\caption{\em Sub-band utilization vs. total number of devices (|A|) with varying traffic intensity ($t_I$) for unconfirmed/confirmed UL transmissions with fixed SF12, $N_{max-UC}$=8, $N_{max-C}$=8, $m_c$=1.}
\label{jpg9}
\end{center}
\end{figure}
\end{itemize}
\subsubsection{Packet Delivery Ratio (PDR)}
Next, we evaluate the PDR for confirmed and unconfirmed frame transmissions in the presence of multiple available channels over the sub-band.
\begin{itemize}
\item \textbf{Fully loaded network ($t_I$=1) with varying maximum transmissions ($N_{max-UC}$ or $N_{max-C}$) }
The results in Fig.~\ref{jpg12}a,b shows the effect on PDR due to the number of network devices (|A|), maximum number of transmissions and the number of available channels $m_c$. As seen in Fig.~\ref{jpg12}a with a single channel gateway (e.g. $m_c$=1) and fully loaded network, the confirmed UL frame transmission gives relatively better results compared to others when the maximum transmissions are 8. However, when the maximum transmissions are limited to $N_{max-C}$=1, the PDR drops because of two reasons; first, because the confirmed transmissions consume more resources (due to ACK transmissions) compared to the unconfirmed ones. Secondly, performing transmissions only one time results in a higher sending rate at the application layer and more unsuccessful UL frames, specifically when there are a lot of devices in the network. Finally, we can see that in a fully loaded network environment, when the maximum transmission limit is high, the choice of unconfirmed/confirmed transmission becomes less important, as they both achieve similar PDR values. 

\begin{figure}[!ht]
\begin{center}
\resizebox{3.4 in}{2.4 in}{\includegraphics{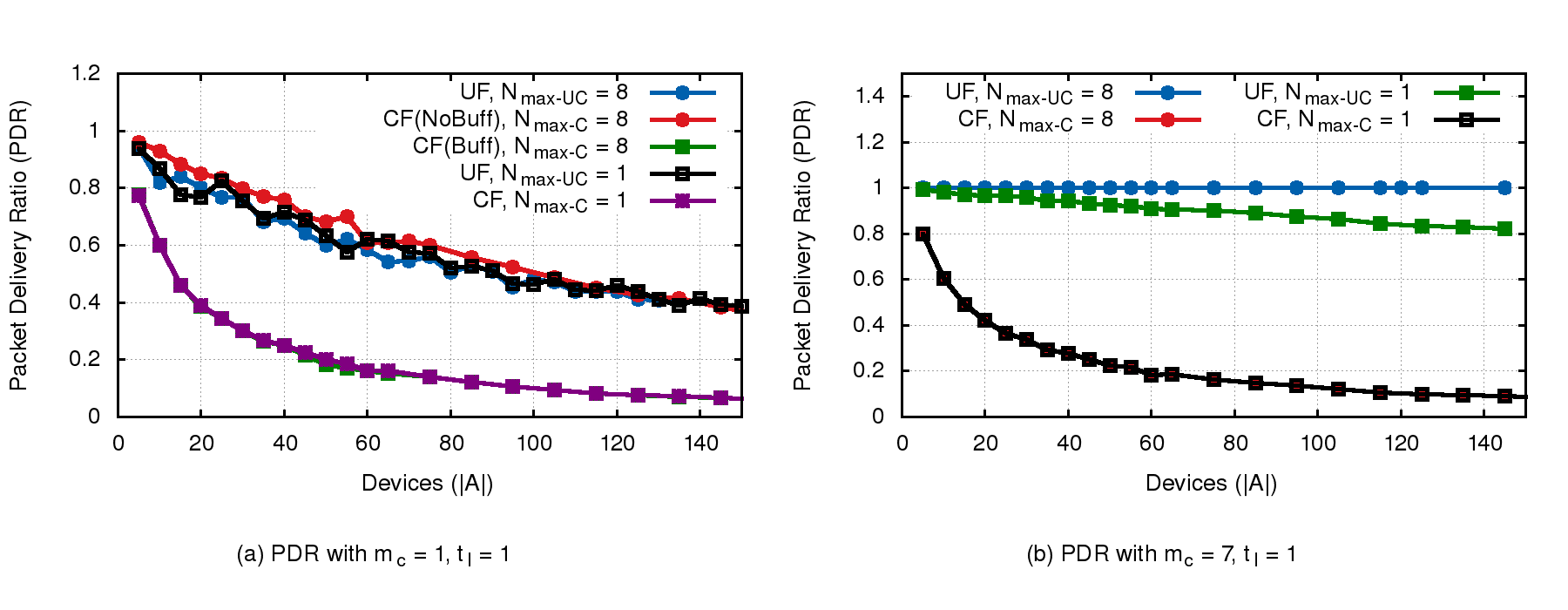}}
\caption{\em PDR vs. network devices (|A|) for confirmed and unconfirmed UL transmissions with different number of maximum transmissions ($N_{max-UC}$ and $N_{max-C}$) and $t_I$=1.}
\label{jpg12}
\end{center}
\end{figure}

Fig.~\ref{jpg12}b shows the results of unconfirmed and confirmed UL transmissions when the number of available channels is increased to 7 ($m_c$=7). Note that with unconfirmed transmissions, when $N_{max-UC}$ is 1 and 8, the application layer PDR is high compared to the case of confirmed UL transmissions. Especially for $N_{max-UC}$=8, the number of frame transmissions increases the chance of successful frame reception, and thus the PDR remains close to 100\% for up to 150 network devices. For the case of only confirmed UL transmissions with max transmission ($N_{max-C}$) set as 8, we notice that the availability of 7 channels decreases the possibility of collisions and allows more new confirmed frames to be sent over the UL successfully. However, due to a higher number of available channels for confirmed transmissions, the ratio of successfully received packets is also high, resulting in a PDR which is the same as in the case of unconfirmed transmissions. 

\item \textbf{Reduced load ($t_I$=0.10) with varying maximum transmissions ($N_{max-UC}$ or $N_{max-C}$) }
Further, we explored the case of a moderate network load, with a traffic intensity of $t_I$=0.10. In Fig.~\ref{jpg13}a the results of a single channel gateway network shows that when the load is high the PDR of confirmed transmissions for both the cases of $N_{max-C}$ 1 and 8 is small compared to the unconfirmed one when $N_{max-UC}$ is 1. Another important point to note is that when the network load is low (e.g. $t_I$=0.10) even a single unconfirmed transmission is enough and achieves a better PDR compared to repetitive UL unconfirmed transmissions of the same frame (e.g. when $N_{max-UC}$ = 8). This is because repetitive unconfirmed transmissions increase the network load and results in more collisions, thereby effectively reducing the PDR, as shown in Fig.~\ref{jpg13}a. From the interesting trend in low load conditions we can conclude that when network devices are less than $\approx$ 40, confirmed UL transmissions with a high maximum transmission limit can be used to achieve better reliability. Comparing this to the earlier results of a single channel gateway network in Fig.~\ref{jpg12}a, an important point to note is that when the network is lightly loaded, repetitive transmissions of the same unconfirmed frames increases the load and results in more collisions. 

In Fig.~\ref{jpg13}b the PDR results based on low traffic intensity ($t_I$=0.10) with all channels available ($m_c$=7) show that when the load on the channel is relatively small, the PDR results follow the patterns for different $N_{max-UC}/N_{max-C}$ and network devices, as expected. In other words, with a higher maximum transmission limit and more available channels, the PDR increases for both confirmed and unconfirmed transmissions.
\begin{figure}[!ht]
\begin{center}
\resizebox{3.4 in}{2.4 in}{\includegraphics{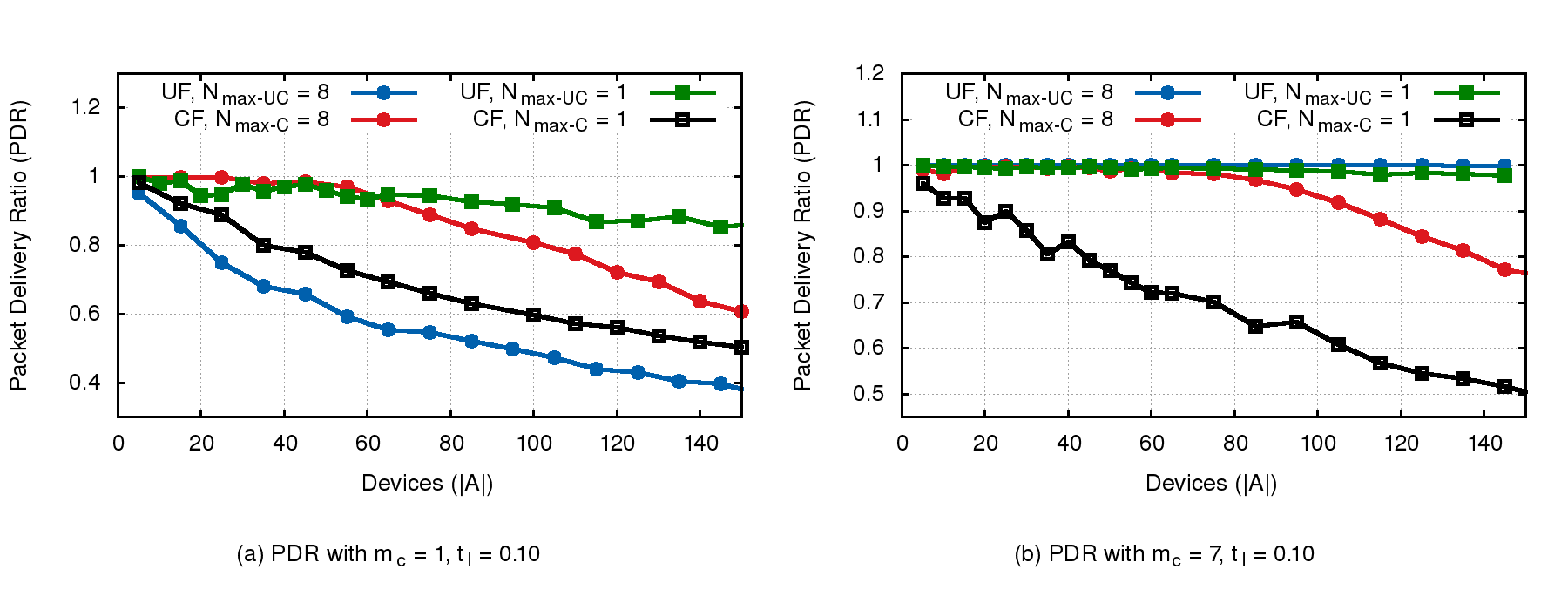}}
\caption{\em PDR vs. network devices (|A|) for confirmed and unconfirmed UL transmissions with different number of maximum transmissions ($N_{max-UC}$ and $N_{max-C}$) and $t_I$=0.10.}
\label{jpg13}
\end{center}
\end{figure}
%

%
%
\end{itemize}
\section{Conclusion}
In this paper, an extensive evaluation of Class A devices in LoRaWAN is performed separately for unconfirmed and confirmed UL transmissions. The results give us valuable new insights on the choice of different parameter values in LoRaWAN affects the overall network performance. For example, in a network with a single device, we found that adaptively changing rate can significantly increase the network performance. With multiple devices in a less loaded network, with a limited number of channels, we found that increasing the number of maximum transmission for unconfirmed frames degrades network performance by reducing the resulting PDR. Also, in such cases we observed that using confirmed transmission is better. Similarly, when the network is highly loaded, we found that unconfirmed transmissions result in better overall performance. The key findings from our work can be used to optimize the LoRaWAN network performance by appropriately choosing different network parameters and device transmission mode based on the network condition.
\appendices


\ifCLASSOPTIONcaptionsoff
  \newpage
\fi

\end{document}